\documentclass[twocolumn,preprintnumbers,amsmath,amssymb]{revtex4}
\usepackage{graphicx}
\usepackage{dcolumn}
\usepackage{bm}

\newcommand{\be}{\begin{equation}}
\newcommand{\bel}[1]{\begin{equation}\label{#1}}
\newcommand{\ee}{\end{equation}}
\newcommand{\bea}{\begin{eqnarray}}
\newcommand{\ba}{\begin{array}}
\newcommand{\eea}{\end{eqnarray}}
\newcommand{\ea}{\end{array}}

\begin{document}

\title{\bf Mechanical filtering in forced-oscillation of two coupled pendulums }

\author{M. Ebrahim Foulaadvand and Davoud Masoumi}

\affiliation{ Department of Physics, Shahid Beheshti (former National) University, P.O. Box 19839-63113, Evin, Tehran, Iran.}

\date{\today}
\begin{abstract}

Forced oscillation of a system composed of two pendulums coupled by a spring in the presence of damping is investigated. In the steady state and
within the small angle approximation we solve the system equations of motion and obtain the amplitudes and phases of in terms of the frequency
of the sinusoidal driving force. The resonance frequencies are obtained and the amplitude ratio is discussed in details. Contrary to a single oscillator, in this two-degree of freedom system four resonant frequencies, which are close to mode frequencies, appear. Within the pass-band interval the system  is shown to exhibit a rich and complicated behaviour. It is shown that damping crucially affects the system properties. Under certain circumstances, the amplitude of the oscillator which is directly connected to the driving force becomes smaller than the one far from it. Particularly we show the existence of a driving frequency at which the connected oscillator's amplitude goes zero.

\end{abstract}

\maketitle
\section{{Introduction}}

Forced oscillation of many body systems constitutes an important and interesting subject in wave phenomena. This subject is normally discussed incompletely in the text books and various aspects have remained unexplored \cite{crawford,marion,pain,french}. In reference \cite{crawford} this issue is discussed in the simplest framework of a system having two degrees of freedom consisting of two pendulums connected to each other by a spring. This system has two mode frequencies $\omega_1$ and $\omega_2>\omega_1$. It is qualitatively argued that if the driving frequency is lower that $\omega_1$ or greater than $\omega_2$ the amplitude of the oscillator directly connected to the driving force is larger than the one far from it. The further the driving frequency from the mode frequency, the larger the amplitude ratio. In this manner the system behaves as a mechanical filter which allows frequencies in the pass band region $[\omega_1,\omega_2]$. Reference [1] does not discuss what occurs in the interesting case where the driving frequency lies in pass-band region and only states that in the pass-band region the amplitudes are comparable to each other. In this paper we aim to quantitatively investigate the properties of such system in the entire range of $\omega \in [0,\infty[$. We show interesting features will arise when $\omega$ lies in the pass-band region. In particular in the absence of damping it is shown there exists an intermediate frequency at which the amplitude of the connected oscillator to the driving force becomes zero.

\section{ Statement of the Problem }

Consider two pendulums $a$ and $b$ each having length $L$ and bob mass $M$ which are coupled via a spring with constant $K$ as illustrated in figure (1). Pendulums are subjected to frictional damping forces and each one has the same damping constant per mass $\Gamma$. Damping can be provided by something rubbing on supporting strings or the bobs. Imagine we derive the system by attaching one of the pendulums say $a$ to a sinusoidal driving force with amplitude per unit length $F_0$ and frequency $\omega$. The equations of motion can easily be written in small angle approximation:

\begin{figure}
\centering
\includegraphics[width=5.5cm]{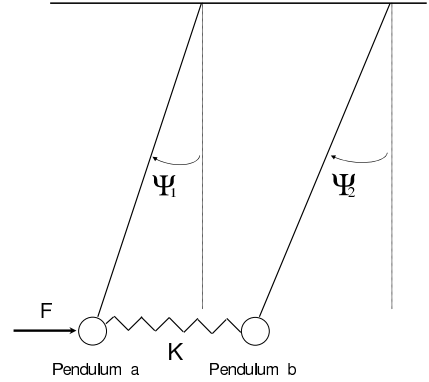}
\caption{ Two coupled pendulums with masses $M$ which are derived by one end by a sinusoidal force. } \label{fig:bz2}
\end{figure}

\be
M\ddot{\Psi}_a=-\frac{Mg}{L}\Psi_a - K(\Psi_a -\Psi_b) -M\Gamma\dot{\Psi}_a + F_0\cos\omega t
\ee

\be
M\ddot{\Psi}_b=-\frac{Mg}{L}\Psi_b + K(\Psi_a -\Psi_b) -M\Gamma\dot{\Psi}_b
\ee

In which $\Psi_a$ and $\Psi_b$ are the angular displacements of pendulums $a$ and $b$. Introducing the normal coordinates $\Psi_1=\frac{1}{2}(\Psi_a+\Psi_a)$ and $\Psi_2=\frac{1}{2}(\Psi_a-\Psi_a)$ and
normal frequencies $\omega_1=\sqrt{\frac{g}{L}}$ and $\omega_2=\sqrt{ \frac{g}{L} + \frac{2K}{M}}$ we recast equations (1) and (2)
into the following decoupled form:

\be
\ddot{\Psi}_1 + \Gamma\dot{\Psi}_1 +\omega^2_1\Psi_1 = \frac{F_0}{2M}\cos\omega t
\ee

\be
\ddot{\Psi}_2 + \Gamma\dot{\Psi}_2 +\omega^2_2\Psi_2 = \frac{F_0}{2M}\cos\omega t
\ee

In steady state when the effect of transients are negligible $\Psi_1$ and $\Psi_2$ appear to be as two independent forced harmonic oscillators with frequency $\omega$:

\be
\Psi_1=A_1(\omega)\cos(\omega t -\delta_1(\omega))
\ee

\be
\Psi_2=A_2(\omega)\cos(\omega t -\delta_2(\omega))
\ee

Amplitudes are obtained as follows \cite{marion}:

\be
A_i(\omega)=\frac{F_0}{2M\sqrt{(\omega^2_i - \omega^2)^2 +\Gamma^2\omega^2 }} ~~~ i=1,2
\ee

Trigonometric functions of the phase differences respect to driving force turn out to be:

\be
\sin\delta_i(\omega)=\frac{\Gamma\omega}{\sqrt{(\omega^2_i - \omega^2)^2 +\Gamma^2\omega^2 }} ~~~ i=1,2
\ee

\be
\cos\delta_i(\omega)=\frac{\omega^2_i -\omega^2}{\sqrt{(\omega^2_i - \omega^2)^2 +\Gamma^2\omega^2 }} ~~~ i=1,2
\ee

\be
\tan\delta_i(\omega)=\frac{\Gamma\omega}{\omega^2_i -\omega^2} ~~~ i=1,2
\ee

The steady state displacements of oscillators $a$ and $b$ will be $\Psi_a=\Psi_1 + \Psi_2$ and $\Psi_b=\Psi_1 - \Psi_2$
respectively. After some straightforward trigonometric manipulations we obtain:

\be
\Psi_a=A_a\cos(\omega t -\delta_a)
\ee

In which the amplitude $A_a$ and phase $\delta_a$ are:

\be A_a=\sqrt{A^2_1 + A^2_2 + 2A_1A_2\cos(\delta_1-\delta_2)}
\ee

\be
\sin\delta_a=\frac{A_1\sin\delta_1 + A_2\sin\delta_2}{A_a}
\ee

\be
\cos\delta_a=\frac{A_1\cos\delta_1 + A_2\cos\delta_2}{A_a}
\ee

\be
\tan\delta_a=\frac{A_1\sin\delta_1 + A_2\sin\delta_2}{A_1\cos\delta_1 + A_2\cos\delta_2}
\ee

In a similar fashion we obtain for oscillator $b$:

\be
\Psi_b=A_b\cos(\omega t -\delta_b)
\ee

\be A_b=\sqrt{A^2_1 + A^2_2 - 2A_1A_2\cos(\delta_1-\delta_2)}
\ee

\be
\sin\delta_b=\frac{A_1\sin\delta_1 - A_2\sin\delta_2}{A_b}
\ee

\be
\cos\delta_b=\frac{A_1\cos\delta_1 - A_2\cos\delta_2}{A_b}
\ee

\be
\tan\delta_b=\frac{A_1\sin\delta_1 - A_2\sin\delta_2}{A_1\cos\delta_1 - A_2\cos\delta_2}
\ee

The main questions we would like to address are as follows:\\

1) What is the dependence of $\frac{A_a}{A_b}$ on the driving frequency $\omega$ ?\\

2) What are the dependence of phases $\delta _a$ and $\delta _b$ on the driving frequency $\omega$ ?\\

We address these questions in two separate cases. In the first case we assume the drag force is absent i.e.; $\Gamma=0$. In the
second case which is far more challenging the drag force is non zero $\Gamma \neq 0$. Let us first discuss case one.

\section{No damping force}

In this case where $\Gamma=0$ the mode amplitudes and phases are simplified as follows:

\be
A_i=\frac{F_0}{2M|\omega^2_i - \omega^2|};~~~~ \sin\delta_i=\tan\delta_i=0
\ee

\be
\cos\delta_i=\frac{\omega^2_i - \omega^2}{|\omega^2_i - \omega^2|} ~~~~ i=1,2
\ee

Three regimes are normally discussed in the literature: $\omega \leq \omega_1<\omega_2$; $\omega_1<\omega \leq \omega_2$ and $\omega_1<\omega_2<\omega$.
We discuss each regime separately.

\subsection{Regime I: $\omega < \omega_1<\omega_2$}

In this regime where $\omega$ is below the pass-band lower cut-off we have: $\cos\delta_1=\cos\delta_2=1$ which give $\delta_1=\delta_2=0$. By (13) we conclude $\sin\delta_a=0$ which implies $\delta_a$ is either zero or $\pi$. The correct choice depends on $\cos\delta_a$. We have
$\cos\delta_a=\frac{A_1 +A_2}{A_a}$. On the other hand $\delta_1=\delta_2=0$ and (12) give $A_a=A_1+A_2$ which in turn gives
rise to $\cos\delta_a=1$. This excludes $\delta_a=\pi$ and therefore we conclude $\delta_a=0$. Similarly we have
$\sin\delta_b=0$ which gives $\delta_b$ is either zero or $\pi$. By (19) we have $\cos\delta_b=\frac{A_1 -A_2}{A_b}$. On the other hand
(17) gives $A_b=|A_1-A_2|$. Having in mind that in regime I we have $\omega \leq \omega_1<\omega_2$ together with (21) we can conclude
that $A_b=A_1-A_2$ which gives rise to $\cos\delta_b=1$. This fixes $\delta_b=0$. We therefore conclude that in regime I both
oscillators $a$ and $b$ are in phase with the driving force. Lastly we evaluate the amplitude ratio $\frac{A_a}{A_b}$.
According to our findings we have

\be
\frac{A_a}{A_b}=\frac{A_1+A_2}{A_1-A_2}>1
\ee

This means oscillator $a$ which is directly connected to driving force has a larger amplitude.

\subsection{Regime II: $\omega_1 \leq \omega \leq \omega_2$}

In this regime we can simply determine $\delta_b$ and $A_b$. According to (22) $\cos\delta_1=-1$ and $\cos\delta_2=1$ which lead to $A_b=A_1 +A_2$. By (19) we have $\cos\delta_b=-1$ which gives $\delta_b=\pi$. However, determination
of $\delta_a$ and $A_a$ is not so simple as that was for $\delta_b$. We note $\cos\delta_1=-1$ and
$\cos\delta_2=1$ give $\delta_1=\pi$ and $\delta_2=0$. Moreover,
$\sin\delta_a=0$ which gives $\delta_a=0$ or $\pi$ and
$\cos\delta_a=\frac{A_2-A_1}{|A_2-A_1|}$. To determine the sign of
the denominator we note that according to (21) if $\omega<\sqrt{\frac{\omega^2_1+\omega^2_2}{2}}$ then $A_1>A_2$ which
fixes $\cos\delta_a=-1$. This yields $\delta_a=\pi$. Since we will frequently encounter the frequency
$\sqrt{\frac{\omega^2_1+\omega^2_2}{2}}$ in the remainder of the paper from now on we denote it by $\tilde{\omega}$.
On the other hand if $\omega>\tilde{\omega}$ we have
$A_1<A_2$ which fixes $\cos\delta_a=1$. This yields $\delta_a=0$. In
the case of equality i.e.; $\omega=\tilde{\omega}$ we have
$A_1=A_2$.\\

To summarize, when the driving frequency lies in pass band that is between the
mode frequencies three distinctive cases are identified:\\

Case 1) ~~~ $\omega_1 \leq \omega<\tilde{\omega}$\\

In this case $\delta_a=\pi$ (oscillator connected to the driving force is out of phase with it !), $\delta_b=\pi$, $A_a=|A_1-A_2|=A_1-A_2$, $A_b=A_1+A_2$. Therefore the amplitude ratio reads:

\be
\frac{A_a}{A_b}=\frac{A_1-A_2}{A_1+A_2}<1.
\ee

Case 2) ~~~ $\tilde{\omega}<\omega \leq \omega_2$\\

In this case $\delta_a=0$ and $A_a=|A_1-A_2|=A_2-A_1$, $A_b=A_1+A_2$. Therefore the amplitude ratio turns out to be:

\be
\frac{A_a}{A_b}=\frac{A_2-A_1}{A_1+A_2}<1.
\ee

Case 3) ~~~ $\tilde{\omega}=\omega$\\

In this case $A_a=0$ that is the connected oscillator to the driving force remains immobile, its phase $\delta_a$ is therefore undetermined.
This case is interesting and introduces a frequency $\tilde{\omega}$ at which the amplitude of the oscillator connected to the driving force is zero while the second oscillator has a non zero amplitude $A_b=2A_1=\frac{F_0}{K}$. Therefore the amplitude ratio $\frac{A_a}{A_b}$ is zero.

\subsection{Regime III: $\omega_1<\omega_2<\omega$}

In this regime the driving frequency is above the pass band upper cut off. We simply obtain $\cos\delta_1=\cos\delta_2=-1$ which give $\delta_1=\delta_2=\pi$. By a similar argument we obtain $\delta_a=\pi$, $\delta_b=0$, $A_a=A_1+A_2$ and $A_b=A_2-A_1$. The amplitude ratio becomes:

\be
\frac{A_a}{A_b}=\frac{A_1+A_2}{A_2-A_1}>1.
\ee

Note that in this case oscillator $a$ is out of phase with the driving force. This aspect is discussed in section 3.3 of reference [1] which qualitatively predicts that when the system is derived above the high cut-off frequency $\omega_2$ its constituents phases resemble to that of the highest mode of free oscillation (out of phase relative to each other). Figure (2) shows the dependence of amplitude ratio on $\omega$ for various values of coupling $K$. Throughout the paper we have used the unit in which $M=g=L=1$. In this unit $\omega_1$ turns out to be one. Moreover, we have set $K=10$ unless otherwise stated. For this value of $K$ we have $\omega_2=4.58 .$ \\

\begin{figure}
\centering
\includegraphics[width=7.5cm]{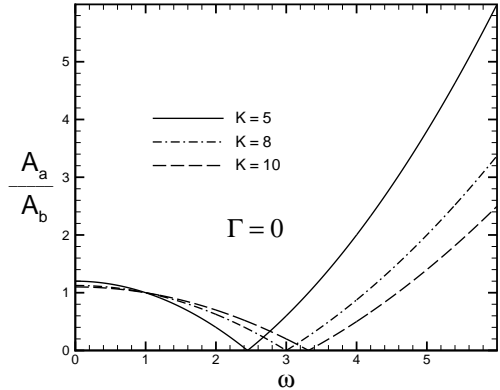}
\caption{ Dependence of ratio $\frac{A_a}{A_b}$ on $\omega$ in absence of damping for three values of coupling parameter $K$. } \label{fig:bz2}
\end{figure}

\section{Effect of damping force}

Now we turn to the more realistic case where damping forces are present
i.e.; $\Gamma \neq 0$. By substitution of (8) and (9) into (15) and (20) we obtain:

$$\tan\delta_a=\frac{\Gamma\omega [(\omega^2_1-\omega^2)^2 +
(\omega^2_2-\omega^2)^2
+2\Gamma^2\omega^2]}{[(\omega^2_1-\omega^2)(\omega^2_2-\omega^2)
+\Gamma^2\omega^2][\omega^2_1 + \omega^2_2-2\omega^2)]} .
$$
\be
\ee
\be
\tan\delta_b=\frac{\Gamma\omega [\omega^2_1 +
\omega^2_2-2\omega^2]}{[(\omega^2_1-\omega^2)(\omega^2_2-\omega^2)
-\Gamma^2\omega^2]} .
\ee

Four independent polynomials appear in the above expressions. Let us
name them as follows:

\be
q(\omega)=(\omega^2_1-\omega^2)(\omega^2_2-\omega^2) +\Gamma^2\omega^2.
\ee

\be
p(\omega)=(\omega^2_1-\omega^2)(\omega^2_2-\omega^2) -\Gamma^2\omega^2.
\ee

\be
h(\omega)=\omega^2_1 + \omega^2_2 -2\omega^2.
\ee

\be
s(\omega)=(\omega^2_1-\omega^2)^2 + (\omega^2_2-\omega^2)^2
+2\Gamma^2\omega^2.
\ee

With these definitions, we have:

\be \tan\delta_a=\frac{\Gamma\omega s}{qh} ~~
\tan\delta_b=\frac{\Gamma\omega h}{p}. \ee

Let us first discuss amplitudes ratio. According to (12) and (17)
$\frac{A_a}{A_b}$ depends on $\cos(\delta_1 - \delta_2)$. If
$\cos(\delta_1 - \delta_2) \geq 0$ then $A_a \geq A_b$ otherwise if
$\cos(\delta_1 - \delta_2)\leq0$ then $A_a \leq A_b$. Using (8) and (9) we
obtain after somewhat algebra:

\be
\cos(\delta_1 - \delta_2)=\frac{q}{(\omega^2_1-\omega^2)^2 +\Gamma^2\omega^2}
\ee

Since the denominator of (34) is positive, the sign of $\cos(\delta_1 -
\delta_2)$ is the same as $q$. The signs of $\delta_a$ and $\delta_b$
depends on signs of $p,q,s$ and $h$. Clearly $s>0$ so we only have
to determine the signs of $p,q$ and $h$ in the range
$0<\omega<\infty$. Polynomial $h$ is quadratic in $\omega$ therefore it would be easy to  determine its sign. By contrast,
$p$ and $q$ are quartic and their sign determination are not trivial. This task is done in details in appendix and the results
are shown in tables (1-7). We only quote and use the results here. Analogous to the case of free damping case we discuss three regimes:

\subsection{Regime I: $\omega < \omega_1<\omega_2$}

According to tables (1-7) when $\omega$ is below the lower cut-off $\omega_1$, $q$ and $h$ are positive so we conclude that $\tan\delta_a>0$ ($\delta_a>0$) which means the oscillator connected to driving force is in phase with it. To determine the sign of $p$, we introduce two frequencies

\be \omega^{(p)}_{\pm}=[\frac{\omega^2_1 + \omega^2_2 + \Gamma^2
\pm \sqrt{\Delta^{(p)}}}{2}]^{\frac{1}{2}} \ee

In which $\Delta^{(p)}$ is defined as follows:

\be \Delta^{(p)}=(\omega^2_1 + \omega^2_2 + \Gamma^2)^2
-4\omega^2_1\omega^2_2 \ee

In appendix we show $\omega^{(p)}_{-}<\omega_1$ and $p>0$ for
$0<\omega<\omega^{(p)}_{-}$ whereas $p<0$ for $\omega^{(p)}_{-}<\omega<\omega_1$
. Therefore if $0<\omega<\omega^{(p)}_{-}$ then $\tan\delta_b>0 \Rightarrow
\delta_b>0$. If $\omega^{(p)}_{-}<\omega<\omega_1$ then $\tan\delta_b<0
\Rightarrow \delta_b<0$. We note that $\omega^{(p)}_{+}>\omega_1$ therefore it
lies outside the region under consideration. Now let us discuss the amplitude ratio
$\frac{A_a}{A_b}$. As shown in tables (3-7) three regimes of $\Gamma$ should be distinguished:\\

$ A:~ 0<\Gamma<\omega_2 -\omega_1$\\

Here have $q>0$ which implies $\cos(\delta_1 - \delta_2)>0$ Therefore the amplitude ratio $\frac{A_a}{A_b}$ is greater than one.\\

$ B:~ \Gamma=\omega_2 -\omega_1$\\

Here have $q<0$ therefore $\cos(\delta_1 - \delta_2)<0$ and therefore
the amplitude ratio $\frac{A_a}{A_b}$ is less than one.\\

$ C:~ \Gamma>\omega_2 -\omega_1$\\

Here $q>0$ therefore $\cos(\delta_1 - \delta_2)>0$ and hence the amplitude ratio $\frac{A_a}{A_b}$ is greater than one.\\

\subsection{Regime II: $\omega_1 \leq \omega \leq \omega_2$}

According to Appendix, sign determination of $q$ crucially depends on $\Gamma$. Let us first discuss $\tan\delta_b$. According to (33) the sign of $\tan\delta_b$ is identical to $\frac{h}{p}$. For $\omega_1<\omega<\tilde{\omega}$ we have $\tan\delta_b<0$ whereas for $\tilde{\omega}<\omega<\omega_2~$ $\tan\delta_b$ becomes positive. Now we discuss $\tan\delta_a$. According to (33) the sign of $\tan\delta_a$
is the same as $qh$. We introduce two frequencies

\be \omega^{(q)}_{\pm}=[\frac{\omega^2_1 + \omega^2_2 - \Gamma^2
\pm \sqrt{\Delta^{(q)}}}{2}]^{\frac{1}{2}} \ee

In which $\Delta^{(q)}$ is defined as follows:

\be \Delta^{(q)}=(\omega^2_1 + \omega^2_2 - \Gamma^2)^2
-4\omega^2_1\omega^2_2 \ee

Five states are identified:\\

$A:~ 0<\Gamma<\frac{\omega^2_2 - \omega^2_1}{\sqrt{2(\omega^2_1 + \omega^2_2)}}$\\

According to appendix we have $\tan\delta_a>0$ for $\omega_1 \leq \omega<\omega^{(q)}_{-}$ and
$\tilde{\omega}<\omega<\omega^{(q)}_{+}$ and $\tan\delta_a<0$ in the remainder of the interval
$[\omega_1,\omega_2]$. Now we discuss amplitude ratio. In the interval $[\omega_1,\omega^{(q)}_{-}[$ and
$]\omega^{(q)}_{+},\omega_2]$ we have $q>0$ which gives $\frac{A_a}{A_b}>1$. In the remainder of the
interval $\frac{A_a}{A_b}<1$. At $\omega=\omega^{(q)}_{\pm}$ we have $\frac{A_a}{A_b}=1$\\

$B: \Gamma=\frac{\omega^2_2 - \omega^2_1}{\sqrt{2(\omega^2_1 + \omega^2_2)}} $\\

According to appendix we have $\tan\delta_a>0$ for $\omega_1 \leq \omega<\omega^{(q)}_{-}$ and $\tan\delta_a<0$
in the remainder of the interval $[\omega_1,\omega_2]$. The amplitude ratio is identical to part A.\\

$C:~ \frac{\omega^2_2 - \omega^2_1}{\sqrt{2(\omega^2_1 + \omega^2_2)}} <\Gamma <\omega_2-\omega_1$\\

In this range of $\Gamma$ $\tan\delta_a>0$ for $\omega_1 \leq \omega<\omega^{(q)}_{-}$ and $\omega^{(q)}_{+}<\omega<\tilde{\omega}$ and
$\tan\delta_a<0$ in the remainder of the interval $[\omega_1,\omega_2]$. The amplitude ratio is identical to that in part A.\\

$D:~ \Gamma = \omega_2-\omega_1$\\

At this particular value of $\Gamma$ $\tan\delta_a>0$ for two sub intervals: $\omega_1 \leq \omega<\omega^{*}$ and $\omega^{*}<\omega<\tilde{\omega}$.
$\tan\delta_a<0$ in the remainder of the interval $[\omega_1,\omega_2]$. Note that in this case, at two special frequencies
$\omega^{*}=\sqrt{\omega_1 \omega_2}$ and ~ $\tilde{\omega}$ the expression $qh$ becomes zero and hence $\tan\delta_a$ goes infinity. The notable point is the amplitude ratio $\frac{A_a}{A_b}$ is greater than one except at $\omega^{*}=\sqrt{\omega_1 \omega_2}$ where it becomes unity. \\

$E:~ \Gamma > \omega_2-\omega_1$\\

In this range of $\Gamma$ $\tan\delta_a>0$ for $\omega_1 \leq \omega<\tilde{\omega}$ and $\tan\delta_a<0$ in the remainder of the interval $[\omega_1,\omega_2]$. The amplitude ratio $\frac{A_a}{A_b}$ is greater than one in the whole range $[\omega_1,\omega_2]$.

\subsection{Regime III: $\omega_1<\omega_2<\omega$}

We now discuss regime III which is much simpler than regime II. In
regime III the driving frequency is above the upper cut-off frequency $\omega_2$. In this
regime $q>0$ and $h<0$. Therefore (33) gives $\delta_a<0$. According to table (2), $\tan\delta_b>0$ for
$\omega_2<\omega<\omega^{(p)}_{+}$ and $\tan\delta_b<0$ for $\omega^{(p)}_{+}<\omega$. Note that at $\omega=\omega^{(p)}_{+}$ we
have $\tan\delta_b=\pm \infty$ depending on how $\omega$ approaches $\omega^{(p)}_{+}$. The amplitude ratio $\frac{A_a}{A_b}$ is
always greater than unity for $\omega_2<\omega$ according to appendix tables.

\subsection{Results and discussion}

In this section we discuss about the role of damping and drive on the oscillators characteristics. By substitution of the modes amplitudes and phases into (12) and (17) we arrive at the following relations for the oscillator amplitudes $A_a$ and $A_b$:

\be
A_a=\frac{F_0}{2M}\sqrt{ \frac{4\omega^4 + \omega_1^4 + \omega_2^4 -4\omega^2(\omega_1^2 + \omega_2^2 -\Gamma^2) +2\omega_1^2\omega_2^2}
{(\omega^2_1 - \omega^2)^2 + \Gamma^2\omega^2)(\omega^2_2 - \omega^2)^2 +\Gamma^2\omega^2) } }
\ee

\be
A_b=\frac{F_0}{2M}\sqrt{ \frac{ \omega_1^4 + \omega_2^4 - 2\omega_1^2\omega_2^2}
{(\omega^2_1 - \omega^2)^2 + \Gamma^2\omega^2)(\omega^2_2 - \omega^2)^2 +\Gamma^2\omega^2) } }
\ee

To find the amplitude resonant frequencies we should set the derivatives of $A_a$ and $A_b$ respect to $\omega$ equal to zero. For $A_a$ this
leads to a polynomial equation of order $10$. For $A_b$ this task leads to a cubic equation for $\omega$ with lengthy coefficients. Despite availability of analytical solution for a cubic equation, we prefer to proceed numerically. Figure (3) depicts the amplitude ratio in the whole range of $\omega$ for various values of $\Gamma$ associated to the states A, B, C , D  and E described in regime II. It is worthwhile mentioning that at $\Gamma=\omega_2 -\omega_1$ the amplitude ratio is an increasing function of $\omega$ whereas in other values of $\Gamma$ there is minimum value for the ratio.\\

\begin{figure}
\centering
\includegraphics[width=7.5cm]{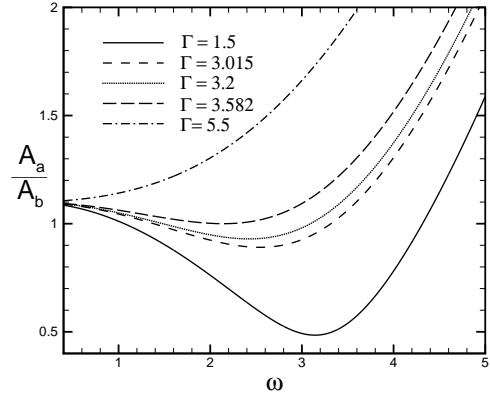}
\caption{ Dependence of ratio $\frac{A_a}{A_b}$ on $\omega$ in presence of damping for some values of damping constant $\Gamma$ in states A to E. } \label{fig:bz2}
\end{figure}

Figure (4) shows the dependence of amplitudes $A_a$ and $A_b$ on $\omega$ for a small damping constant. There exist two peaks at which both amplitudes are maximised. The peak values of $\omega$ are resonant frequencies and are very close to $\omega_1$ and $\omega_2$. By increasing $\Gamma$ the resonant frequencies of oscillators $a$ and $b$ begin to deviate from each other. In fact, each oscillator posses its own resonant frequencies. To see this more explicitly in figure (5) we have drawn amplitude dependence on $\omega$ for larger values of $\Gamma$. Separation of resonant frequencies is apparent.

\begin{figure}
\centering
\includegraphics[width=7.5cm]{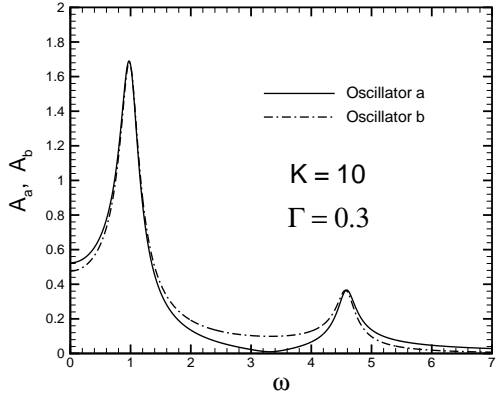}
\caption{ Dependence of amplitudes $A_a$ and $A_b$ on $\omega$ for a small value of damping constant $\Gamma=0.3$. } \label{fig:bz2}
\end{figure}

\begin{figure}
\centering
\includegraphics[width=7.5cm]{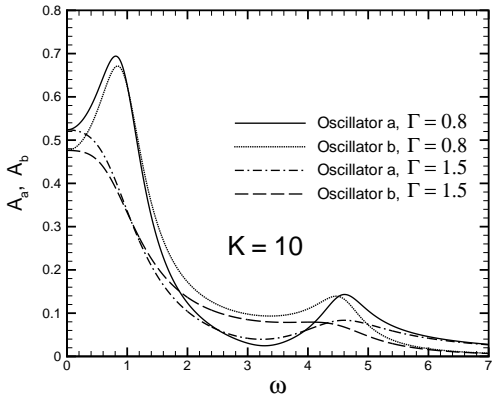}
\caption{ Dependence of amplitudes $A_a$ and $A_b$ on $\omega$ for larger values of damping constant $\Gamma$. } \label{fig:bz2}
\end{figure}

Figure (6) exhibits $\tan \delta_b$ in the whole range of $\omega$ for two values of $\Gamma$.

\begin{figure}
\centering
\includegraphics[width=7.5cm]{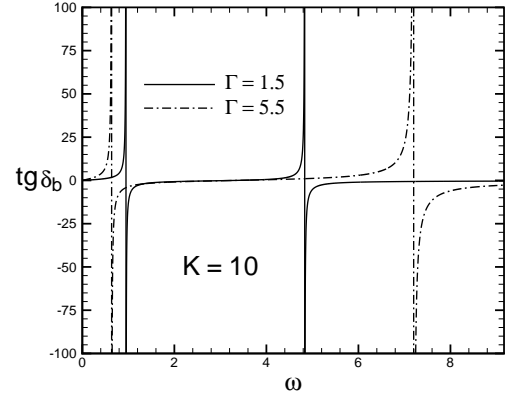}
\caption{ Dependence of $\tan \delta_b$ on $\omega$ for two values of damping constant $\Gamma$. } \label{fig:bz2}
\end{figure}

Notice that $\tan \delta_b$ exhibits two jumps (discontinuities) for each value of $\Gamma$. These jumps occur at roots of $p(\omega)$ which are $\omega^{(p)}_{\pm}$. In figures (7-11) we have shown $\tan \delta_a$ vs $\omega$ for those values of $\Gamma$ in figure (3).

\begin{figure}
\centering
\includegraphics[width=7.5cm]{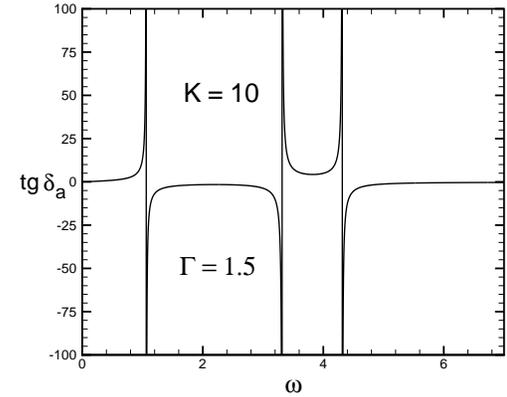}
\caption{ Dependence of $\tan \delta_a$ on $\omega$ for damping constant $\Gamma=1.5$ in state A. } \label{fig:bz2}
\end{figure}

\begin{figure}
\centering
\includegraphics[width=7.5cm]{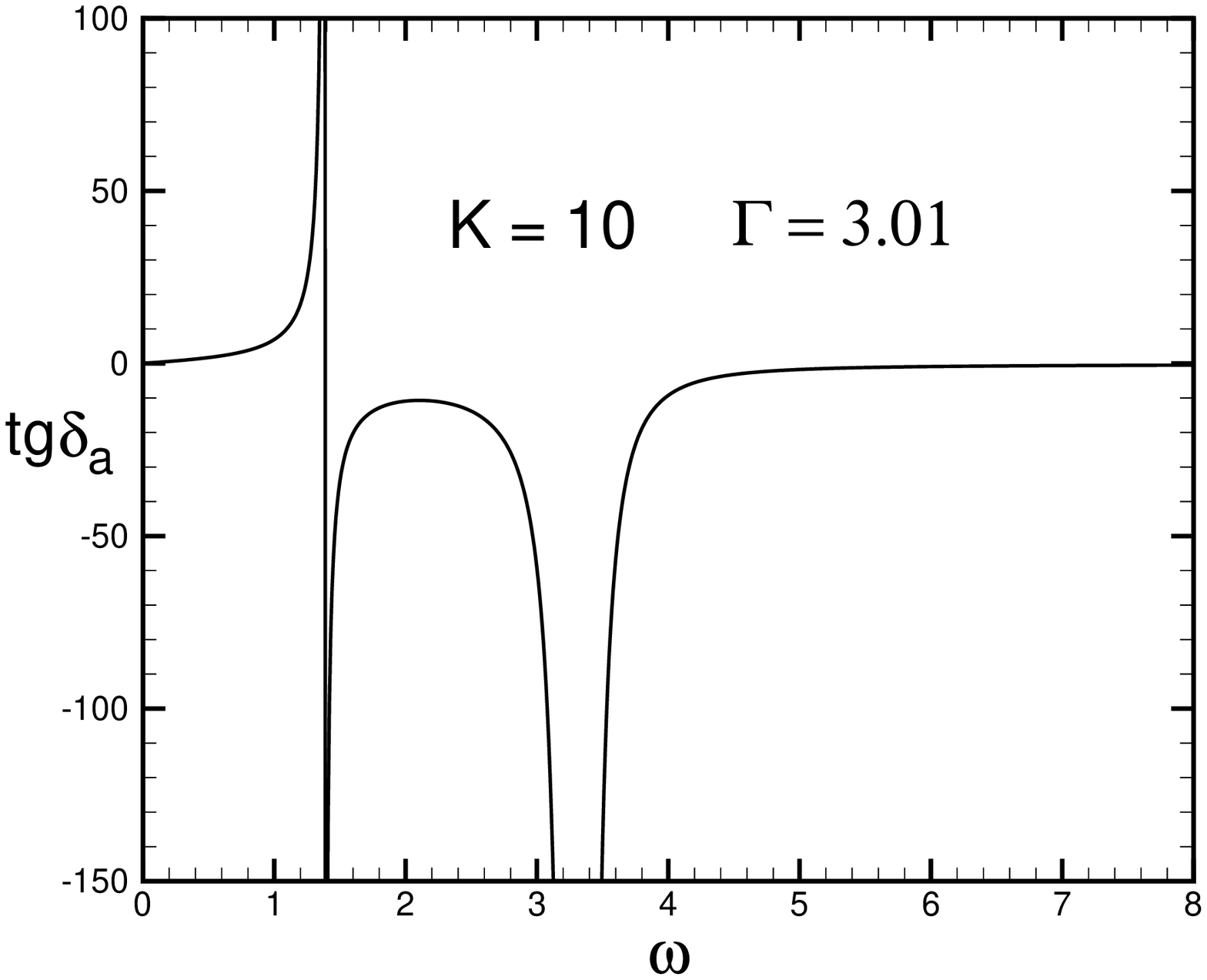}
\caption{ Dependence of $\tan \delta_a$ on $\omega$ for damping constant $\Gamma=3.015$ in state B. } \label{fig:bz2}
\end{figure}

\begin{figure}
\centering
\includegraphics[width=7.5cm]{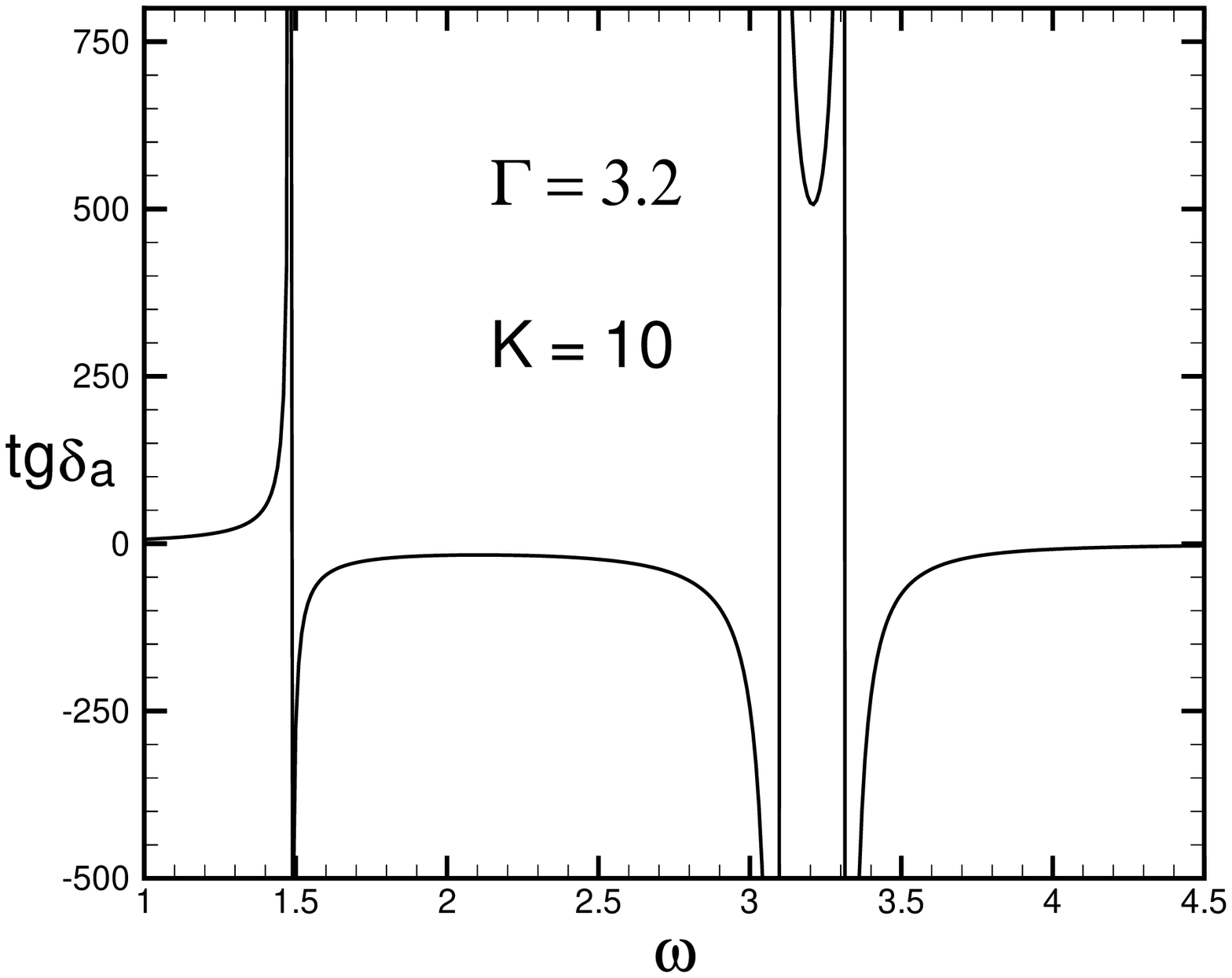}
\caption{ Dependence of $\tan \delta_a$ on $\omega$ for damping constant $\Gamma=3.2$ in state C. } \label{fig:bz2}
\end{figure}

\begin{figure}
\centering
\includegraphics[width=7.5cm]{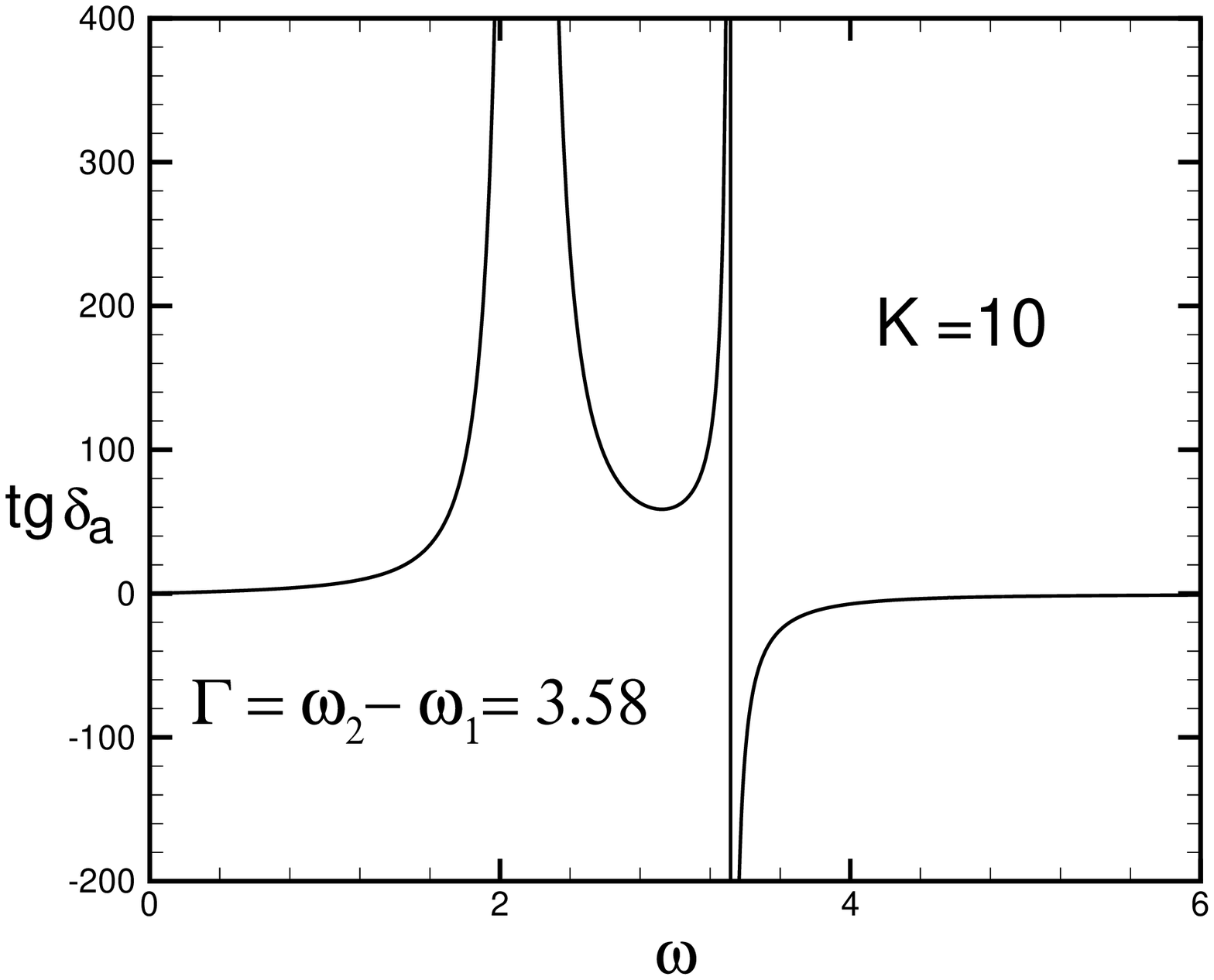}
\caption{ Dependence of $\tan \delta_a$ on $\omega$ for damping constant $\Gamma=3.58$ in state D. } \label{fig:bz2}
\end{figure}

\begin{figure}
\centering
\includegraphics[width=7.5cm]{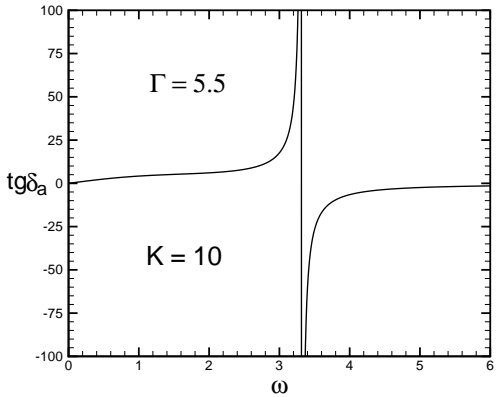}
\caption{ Dependence of $\tan \delta_a$ on $\omega$ for damping constant $\Gamma=5.5$ in state E. } \label{fig:bz2}
\end{figure}

In contrast to $\tan \delta_b$ the number of jumps in $\tan \delta_a$ depends on the value of $\Gamma$. We can have one, two and three jumps in figures (7-11). We always have a jump at $\omega=\tilde{\omega}$ which is the root of $h(\omega)$. Other jumps are associated to $\omega^{(q)}_{\pm}$(roots of $q(\omega)$) which their existence depend on the value of $\Gamma$.

\section{acknowledgement}

We appreciate N. Moghni'odolleh, A. Saffar Zadeh and A. Langari for their useful help.

\section{Appendix}

In this appendix we determine the sign of polynomials $h,p$ and $q$ in variation of $\omega$. Since $h$ is quadratic in $\omega$ its sign can simply be determined. For later purposes we need to specify the location of $\tilde{\omega}=\sqrt{\frac{\omega^2_1 + \omega^2_2}{2}}$ respect to $\omega_1$ and $\omega_2$. According to the following inequality we arrive at the following sign table for $h$:\\

\be
\omega_1<\tilde{\omega}<\omega_2
\ee

\be
\begin{tabular}{ccccc}
  $\omega$&0&$\omega_1$&$\tilde{\omega}$&$\omega_2$ \\
  \hline
  $h$ & ~~~~~~~~~~~~~~~~+ & ~~~~~~~~~~~~~~~+ & ~~~~~~~~~~~~~~~- & ~~~~~~~~~~~~~~~- \\
  \end{tabular}\\
\ee

\subsection{Sign determination of $p$}

Introducing $\Omega=\omega^2$~ polynomial $p$ can be written as:\\

\be
p=\Omega^2 -(\omega^2_1 + \omega^2_2 + \Gamma^2)\Omega + \omega^2_1 \omega^2_2
\ee

Eq. (43) is now quadratic in $\Omega$. Its discriminant $\Delta^{(p)}$ turns out to be: $\Delta^{(p)}=(\omega^2_1 + \omega^2_2 + \Gamma^2)^2 -4\omega^2_1 \omega^2_2$. Utilizing the inequality $(\omega_1 - \omega_2)^2 + \Gamma^2>0$ we can simply show that $\Delta^{(p)}>0$. This implies that equation (43) posses two real solutions:

\be
\Omega^{(p)}_{\pm}=\frac{\omega^2_1 + \omega^2_2 + \Gamma^2 \pm \sqrt{\Delta^{(p)}}}{2}
\ee

On the other hand we have : $\omega^2_1 + \omega^2_2 + \Gamma^2 > \sqrt{\Delta^{(p)}}$ which gives rise to $\Omega^{(p)}_{-}>0$. Therefore we can introduce two frequencies:

\be
\omega^{(p)}_{\pm}=\sqrt{ \Omega^{(p)}_{\pm} }
\ee

Now the sign of $p$ is easy to determine: $p$ is negative if $\omega^{(p)}_{-}<\omega<\omega^{(p)}_{+}$ and positive elsewhere. At
$\omega^{(p)}_{\pm}$ polynomial $p$ becomes zero. Next we should determine the positions of $\omega_1$ and $\omega_2$ respect to $\omega^{(p)}_{\pm}$. We claim:

\be
\omega^{(p)}_{-}<\omega_1.
\ee

The proof is quite simple: squaring both sides of (46) and using the definition of $\omega^{(p)}_{-}$ we arrive at the inequality $(\omega^2_2 - \omega^2_1 + \Gamma^2)<\sqrt{\Delta}$. Squaring both sides of this inequality after some straightforward algebra we reach to a true statement $\Gamma^2>0$ which completes the proof. Similarly we can show that:

\be
\omega^{(p)}_{+}>\omega_1.
\ee

In a rather similar fashion we can prove:

\be
\omega^{(p)}_{+}>\omega_2.
\ee

The following table depicts the sign determination of polynomial $p$. For the sign of $\delta_b$ we have to determine the sign of $\frac{h}{p}$ hence we simultaneously exhibit the signs of $h$ and $\frac{h}{p}$ in the following table.

\be
\begin{tabular}{ccccccc}
  $\omega$ & 0 & $\omega^{(p)}_{-}$ & $\omega_1$ & $\tilde{\omega}$ & $\omega_2$ & $\omega^{(p)}_{+}$ \\
  \hline
  $h$ & ~~~~~~~~+ & ~~~~~~~+ & ~~~~~~~+ & ~~~~~~~- & ~~~~~~~~- & ~~~~~~~- \\
  \hline
  $p$ & ~~~~~~~~+ & ~~~~~~~- & ~~~~~~~- & ~~~~~~~- & ~~~~~~~~- & ~~~~~~~+ \\
  \hline
  $h/p$& ~~~~~~~~+ & ~~~~~~~- & ~~~~~~~- & ~~~~~~~+ & ~~~~~~~~+ & ~~~~~~~- \\  
\end{tabular}\\
\ee

\subsection{Sign determination of $q$}

Introducing $\Omega=\omega^2$ polynomial $q(\omega)$ can be cast into the following form:

\be
q=\Omega^2 -(\omega^2_1 + \omega^2_2 - \Gamma^2)\Omega + \omega^2_1 \omega^2_2
\ee

In terms of $\Omega$ polynomial $q$ becomes quadratic with discriminant $\Delta^{(q)}=(\omega^2_1 + \omega^2_2 - \Gamma^2)^2 -4\omega^2_1 \omega^2_2$.
Contrary to polynomial $p$ where its discriminant was shown to be positive, here $\Delta^{(q)}$ can take any sign which complicates the problem. Therefore we should determine the sign of $\Delta^{(q)}$ {\it in priori}. Three distinctive cases arise:\\

A)~ $\Delta^{(q)}>0$:\\

In this case we have two real roots:

\be
\Omega^{(q)}_{\pm}=\frac{\omega^2_1 + \omega^2_2 - \Gamma^2 \pm \sqrt{\Delta^{(q)}}}{2}
\ee

Let us find out the condition under which $\Delta^{(q)}>0$. Using the definition of $\Delta^{(q)}$, it can be easily shown that the sufficient condition for positivity of $\Delta^{(q)}$ is:

\be
|\omega^2_1 + \omega^2_2 - \Gamma^2|>2 \omega_1 \omega_2
\ee

If $\omega^2_1 + \omega^2_2 - \Gamma^2>0$ i.e.; if $0<\Gamma<\sqrt{\omega^2_1 + \omega^2_2}$ condition (52) reduces to

\be
\Gamma<\omega_2 - \omega_1
\ee

If $\omega^2_1 + \omega^2_2 - \Gamma^2<0$ i.e., if $\sqrt{\omega^2_1 + \omega^2_2}<\Gamma$ condition (52) reduces to

\be
\Gamma>\omega_2 + \omega_1
\ee

B)~ $\Delta^{(q)}=0$:\\

In this case we have a degenerate real root:

\be
\Omega^{*}=\frac{\omega^2_1 + \omega^2_2 - \Gamma^2}{2}
\ee

Putting $\Delta^{(q)}$ equal to zero we should solve the following equation to find out the condition.

\be
|\omega^2_1 + \omega^2_2 - \Gamma^2|=2\omega_1 \omega_2
\ee

If $\omega^2_1 + \omega^2_2 - \Gamma^2 \geq 0 ~(0<\Gamma \leq \sqrt{\omega^2_1 + \omega^2_2})$ (56) gives $\Gamma=\omega_2 - \omega_1$ whereas if
$\omega^2_1 + \omega^2_2 - \Gamma^2 \leq 0~(\sqrt{\omega^2_1 + \omega^2_2} \leq \Gamma)$ we have $\Gamma=\omega_2 + \omega_1$.\\

C)~ $\Delta^{(q)}<0$:\\

In this case (50) has no real root. When $\Delta^{(q)}<0$ we have  $|\omega^2_1 + \omega^2_2 - \Gamma^2|<2\omega_1 \omega_2$. It can simply be shown that this condition is equivalent to $\omega_2-\omega_1<\Gamma<\omega_2+\omega_1$.\\

Now we can determine the sign of $\Delta^{(q)}$ in the entire range of $0<\Gamma<\infty$. The following table exhibits the sign of $\Delta^{(q)}$. In this table we also show the sign of the expression $f(\Gamma)=\omega^2_1 + \omega^2_2 - \Gamma^2$ which appears in (51).

\be
\begin{tabular}{ccccc}
  $\Gamma$ & 0 & $\omega_2 - \omega_1$ & $\sqrt{2} \tilde{\omega}$  & $\omega_1 + \omega_2$   \\
  \hline
  $\Delta^{(q)}$ & ~~~~~~~~~~~+ & ~~~~~~~~~~~~~~~- & ~~~~~~~~~~~~~~- & ~~~~~~~~~~~+   \\
  \hline
  $f(\Gamma)$ & ~~~~~~~~~~~+ & ~~~~~~~~~~~~~~~+ & ~~~~~~~~~~~~~~- & ~~~~~~~~~~~-   \\
  \hline

\end{tabular}\\
\ee

Note that in the interval $\Gamma \in ]\omega_2-\omega_1,\omega_2+\omega_1[$ within which $\Delta^{(q)}<0$ the sign of $q(\omega)$ coincides with the sign of the highest coefficient in (50) (positive). According to the above table for $0<\Gamma<\omega_2-\omega_1$ we have $\Delta^{(q)}>0$ hence both $\Omega^{(q)}_{+}$ and $\Omega^{(q)}_{-}$ are positive (by the very definition of $\Delta^{(q)}$) and we can define:

\be
\omega^{(q)}_{\pm}=\sqrt{\Omega^{(q)}_{\pm}}.
\ee

By contrast, in the interval $\Gamma> \omega_2+\omega_1$ both $\Omega^{(q)}_{+}$ and $\Omega^{(q)}_{-}$ are negative. This means that forth order polynomial $q$ (in terms of $\omega$) has no real root therefore it is either positive or negative. Since the coefficient of the highest term is positive we conclude that $q$ is positive in the entire range $0<\omega<\infty$. The points $\Gamma=\omega_2-\omega_1$ and $\Gamma=\omega_2+\omega_1$ need special attention. At $\Gamma=\omega_2-\omega_1$ we have $\Omega^{*}=\frac{\omega^2_1 + \omega^2_2 - \Gamma^2}{2}=\omega_1\omega_2>0$ therefore we conclude that $q(\omega)=(\omega^2-\Omega^{*})^2=(\omega^2-\omega_1\omega_2)^2\geq0$. At the second special point $\Gamma=\omega_2+\omega_1$ we have $\Omega^{*}=\frac{\omega^2_1 + \omega^2_2 - \Gamma^2}{2}=-\omega_1\omega_2<0$ therefore $q(\omega)=(\omega^2+\omega_1\omega_2)^2>0$.\\

The next stage is to determine the locations of $\omega_1$ and $\omega_2$ respect to $\omega^{(q)}_{\pm}$ and $\omega^{*}=\sqrt{\Omega^{*}}=
\sqrt{\omega_1\omega_2}$. It is sufficient to consider two cases: $0<\Gamma<\omega_2 - \omega_1$ and $\Gamma=\omega_2 - \omega_1$ since in the remaining range of $\Gamma$ there is no real root $\omega$ for $q$.\\

A:~ $0<\Gamma<\omega_2 - \omega_1$\\

In this range of $\Gamma$ we claim $\omega^{(q)}_{-}>\omega_1$. To show this we square both sides and use the definition of $\omega^{(q)}_{-}$ in (45) to reach this inequality:

\be
\omega^2_2 - \omega^2_1 -\Gamma^2<\sqrt{\Delta^{(q)}}.
\ee

It is not difficult to show that the left hand side of (59) is positive for the range $0<\Gamma<\omega_2 - \omega_1$. Denoting the left hand side by $\gamma$ we have $\gamma>0$ for $0<\Gamma<\sqrt{\omega^2_2 - \omega^2_1}$. On the other hand we have $\omega_2 - \omega_1<\sqrt{\omega^2_2 - \omega^2_1}$ therefore $\gamma>0$ in the range $0<\Gamma<\omega_2 - \omega_1$. Now we can square both sides of (59) without reversing the inequality sign. After some straight forward algebra we arrive to a true statement $\Gamma^2>0$ which proves our claim. In a similar fashion we can show that $\omega^{(q)}_-<\omega_2$. Now we turn to $\omega^{(q)}_{+}$. Since $\omega^{(q)}_{-}<\omega^{(q)}_{+}$ and $\omega^{(q)}_{-}>\omega_1$ we conclude that $\omega^{(q)}_{+}>\omega_1$. We claim that $\omega^{(q)}_{+}<\omega_2$. The argument is similar to those for $\omega^{(q)}_{-}$ and we omit the proof. \\

B:~ $\Gamma=\omega_2 - \omega_1$\\

Substituting this value of $\Gamma$ into (55) gives $\Omega^{*}=\omega_1\omega_2$ which yields $\omega^*=\sqrt{\omega_1 \omega_2}$. This means $\omega^*$ is the geometrical  average of mode frequencies therefore we have $\omega_1<\omega^*<\omega_2$. Subsequent to our previous discussions, for $\Gamma>\omega_2 - \omega_1$ we have $q>0$.\\

We are not yet at the end of the story. The point is that the values of $\omega$ at which $q$, $p$ and $h$ change sign do not coincide. This makes the problem complicated. we recall that to find the signs of $\delta_a$ and $\delta_b$ we have to determine the sign of $qh$ and $\frac{h}{p}$ respectively. For this purpose, we should determine the location of $\tilde{\omega}=\sqrt{\frac{\omega^2_1 + \omega^2_2}{2}}$ respect to $\omega^{(p)}_{\pm},\omega^{(q)}_{\pm}$ and $\omega^*$ correspondingly. Now we discuss the sign of $qh$ which gives the sign of $tg\delta_a$. As there are three regimes for $\Gamma$ we discuss them separately. In the first regime i.e.; $0<\Gamma<\omega_2 - \omega_1$ we claim $\omega^{(q)}_-<\tilde{\omega}$. This is evident by the definition of $\omega^{(q)}_-$ as in (51,58). To determine the location of $\tilde{\omega}$ respect to $\omega^{(q)}_+$ we rewrite $\omega^{(q)}_+$ as follows which is more suitable for comparison:

\be
\omega^{(q)}_+=\sqrt{ \tilde{\omega}^2 +\frac{\sqrt{\Delta^{(q)}} - \Gamma^2}{2} }
\ee

Clearly if $\sqrt{\Delta^{(q)}} > \Gamma^2$ we have $\omega^{(q)}_+ >\tilde{\omega}$ and vice versa. In the case of equality i.e.;
$\sqrt{\Delta^{(q)}} = \Gamma^2$ then $\omega^{(q)}_+ = \tilde{\omega}$ which gives (by definition of $\Delta^{(q)}$) :

\be
\Gamma = \frac{\omega^2_2 - \omega^2_1}{\sqrt{2(\omega^2_1 + \omega^2_2)}}
\ee

The condition $\sqrt{\Delta^{(q)}} > \Gamma^2$ implies:

\be
\Gamma<\frac{\omega^2_2 - \omega^2_1}{\sqrt{2(\omega^2_1 + \omega^2_2)}}
\ee

Similarly the condition $\sqrt{\Delta^{(q)}} < \Gamma^2 $ implies :

\be
\Gamma>\frac{\omega^2_2 - \omega^2_1}{\sqrt{2(\omega^2_1 + \omega^2_2)}}
\ee

It is straightforward to show $\frac{\omega^2_2 - \omega^2_1}{\sqrt{2(\omega^2_1 + \omega^2_2)}}<\omega_2 - \omega_1$.

At the specific point $\Gamma=\omega_2 - \omega_1$ we can easily show that $\omega^*=\sqrt{\omega_1\omega_2}<\tilde{\omega}.$
We are now able to exhibit the simultaneous sign tables of $q$, $h$ and hence $tg\delta_a$.\\

I:~ $0<\Gamma<\frac{\omega^2_2 - \omega^2_1}{\sqrt{2(\omega^2_1 + \omega^2_2)}}$

\be
\begin{tabular}{ccccccc}
  $\omega$ & 0 & $\omega_1$ & $\omega^{(q)}_-$ & $\tilde{\omega}$ & $\omega^{(q)}_+$ & $\omega_2$ \\
  \hline
  $q$ & ~~~~~~~~~~+ & ~~~~~~~~~+ & ~~~~~~~~~- & ~~~~~~~~~- & ~~~~~~~~~~+ & ~~~~~~+ \\
  \hline
  $h$ & ~~~~~~~~~~+ & ~~~~~~~~~+ & ~~~~~~~~~+ & ~~~~~~~~~- & ~~~~~~~~~~- & ~~~~~~- \\
  \hline
  $(qh)^{-1}$& ~~~~~~~~~~+ & ~~~~~~~~~+ & ~~~~~~~~~- & ~~~~~~~~~+ & ~~~~~~~~~~- & ~~~~~~- \\  
\end{tabular}
\ee

II: $\Gamma = \frac{\omega^2_2 - \omega^2_1}{\sqrt{2(\omega^2_1 + \omega^2_2)}}$

\be
\begin{tabular}{cccccc}
  $\omega$ & 0 & $\omega_1$ & $\omega^{(q)}_-$ &  $\tilde{\omega}=\omega^{(q)}_+$ & $\omega_2$ \\
  \hline
  $q$ & ~~~~~~~~~~+ & ~~~~~~~~~+ & ~~~~~~~~~- & ~~~~~~~~~~~~~~~+ & ~~~~~~+ \\
  \hline
  $h$ & ~~~~~~~~~~+ & ~~~~~~~~~+ & ~~~~~~~~~+ & ~~~~~~~~~~~~~~~- & ~~~~~~- \\
  \hline
  $(qh)^{-1}$& ~~~~~~~~~~+ & ~~~~~~~~~+ & ~~~~~~~~~- & ~~~~~~~~~~~~~~~- & ~~~~~~- \\  
\end{tabular}
\ee

III:~$\frac{\omega^2_2 - \omega^2_1}{\sqrt{2(\omega^2_1 + \omega^2_2)}}<\Gamma<\omega_2 -\omega_1$

\be
\begin{tabular}{ccccccc}
  $\omega$ & 0 & $\omega_1$ & $\omega^{(q)}_-$ & $\omega^{(q)}_+$ & $\tilde{\omega}$ & $\omega_2$ \\
  \hline
  $q$ & ~~~~~~~~~~+ & ~~~~~~~~~+ & ~~~~~~~~~- & ~~~~~~~~~+ & ~~~~~~~~~~+ & ~~~~~~+ \\
  \hline
  $h$ & ~~~~~~~~~~+ & ~~~~~~~~~+ & ~~~~~~~~~+ & ~~~~~~~~~+ & ~~~~~~~~~~- & ~~~~~~- \\
  \hline
  $(qh)^{-1}$& ~~~~~~~~~~+ & ~~~~~~~~~+ & ~~~~~~~~~- & ~~~~~~~~~+ & ~~~~~~~~~~- & ~~~~~~- \\  
\end{tabular}
\ee

IV:~$\Gamma=\omega_2 -\omega_1$

\be
\begin{tabular}{cccccc}
  $\omega$ & 0 & $\omega_1$ & $\omega^{*}$ & $\tilde{\omega}$ & $\omega_2$ \\
  \hline
  $q$ & ~~~~~~~~~~+ & ~~~~~~~~~+ & ~~~~~~~~~+ & ~~~~~~~~~~+ & ~~~~~~~~~+ \\
  \hline
  $h$ & ~~~~~~~~~~+ & ~~~~~~~~~+ & ~~~~~~~~~+ & ~~~~~~~~~~- & ~~~~~~~~~- \\
  \hline
  $(qh)^{-1}$& ~~~~~~~~~~+ & ~~~~~~~~~+ & ~~~~~~~~~+ & ~~~~~~~~~~- & ~~~~~~~~~- \\  
\end{tabular}
\ee

V:~$\Gamma>\omega_2 -\omega_1$

\be
\begin{tabular}{ccccc}
  $\omega$ & 0 & $\omega_1$ & $\tilde{\omega}$ & $\omega_2$ \\
  \hline
  $q$ & ~~~~~~~~~~+ & ~~~~~~~~~+ & ~~~~~~~~~~+ & ~~~~~~~~~+ \\
  \hline
  $h$ & ~~~~~~~~~~+ & ~~~~~~~~~+ & ~~~~~~~~~~- & ~~~~~~~~~- \\
  \hline
  $(qh)^{-1}$& ~~~~~~~~~~+ & ~~~~~~~~~+ & ~~~~~~~~~~- & ~~~~~~~~~- \\  
\end{tabular}
\ee

\end{document}